\documentstyle[12pt]{article}
\begin{document}
\title{NN, N$\Delta$ Couplings and the Quark Model\thanks{Research
supported in part by the National Science Foundation and by the Department of
Energy }}

\author{Thomas R. Hemmert$^a$, Barry R. Holstein$^a$ and
        Nimai C. Mukhopadhyay$^b$ \\
        $^a$Department of Physics and Astronomy \\
        University of Massachusetts \\
        Amherst, MA 01003 \\
        $^b$Physics Department \\
        Rensselaer Polytechnic Institute \\
        Troy, NY 12180 - 3590}

\maketitle

\begin{abstract}
We examine  mass-corrected SU(6) symmetry
predictions in the quark model relating
vector, axial-vector and strong NN and N$\Delta$ coupling, and demonstrate
that  the experimental N$\Delta$ value is significantly higher
than predicted in each case. Nevertheless the Goldberger-Treiman
relation is satisfied
in both sectors. Possible origins of the discrepancy of the quark model
predictions with experiments are discussed.
\end{abstract}

\section{Introduction}
Particle physicists generally think of the nucleon and the $\Delta$(1232)
baryon as being closely related partners, {\it i.e.} the internal
quark dynamics is assumed to be identical with the mass difference arising from
the effects of
color-hyperfine interactions. Indeed the simple constituent quark wavefunctions
display an explicit SU(6) symmetry whereby the nucleon and the delta share a
56-dimensional representation.\cite{1} Within such a model one can calculate
the vector and axial form factors $f_{i}$ and $g_{i}$ as well as the strong
coupling constant $g_{\pi NN}$ for the pion-nucleon system at low
momentum transfer. Likewise one can evaluate the corresponding
nucleon-delta vector and axial vector form factors
$c_{i}$ and $d_{i}$, as well as the strong coupling
constant $g_{\pi N\Delta}$, at low $q^{2}$. The overall scale depends,
however, on {\it a priori} unknown quark wave functions
of the constituent quarks.
By combining the results of the nucleon and delta calculations, one can
eliminate this wavefunction dependence and obtain definite predictions
relating corresponding NN and N$\Delta$ quantities. As we shall
demonstrate, the experimental N$\Delta$ amplitudes are found to be
systematically larger than predicted in each case, and the origin of this
effect is unclear. Nevertheless, in both NN and N$\Delta$ sectors, we find that
the Goldberger-Treiman \cite{2} relation, required by chiral invariance, is
valid.

In the succeeding sections,  we analyze in turn the vector and axial form
factors of the NN and N$\Delta$ systems, as well as the strong $\pi NN$ and
$\pi N\Delta$ coupling constants. After examining the validity of the
NN and $N\Delta$ Goldberger-Treiman relations, we close our paper with a
summary and some speculations concerning these quark model discrepancies.

\section{Quark Model Calculations}

\subsection{Vector Form Factors}
We begin our discussion with the charge-changing polar vector transition
between neutron and proton, for which the most general matrix element can
be written from spin-parity considerations, in terms of three structure
functions $f_{i}(q^{2})$,
\begin{eqnarray}
J_{\mu}^{\it V}({\bf p},{\bf p'}) & = & \langle p({\bf p'})|
                                             J_{\mu}^{\it V}|n({\bf p})\rangle
                                             \nonumber \\
                                  & = & \bar{u}({\bf p'})
\left\{ f_{1}(q^{2}) \gamma_{\mu} + \frac{i f_{2}(q^{2})}{2M} \sigma_{\mu\nu}
q^{\nu} + \frac{f_{3}(q^{2})}{2M} q_{\mu} \right\} u({\bf p}).
\label{eq:veccur1}
\end{eqnarray}
Here $u, \bar{u}$ are plane-wave nucleon spinors, $M$ is the nucleon mass and
$q_{\mu}=p'_{\mu}-p_{\mu}$ is the four-momentum transfer.
Our goal is to determine the size of these three form factors at low $q^{2}$
using the constituent quark model. This task is made easier by use of the
conserved vector current (CVC) hypothesis, which requires\cite{3}
\begin{equation}
f_{1}(0)=1, \hspace{5 mm} \mbox{and} \hspace{5 mm} f_{3}(q^{2})=0,
\end{equation}
and is consistent with quark model considerations. In order to confront
momentum space expressions such as Eq.~\ref{eq:veccur1} with coordinate
space quark wavefunction calculations, we shall use the wavepacket (WP)
formalism, which utilizes the function $\varphi ({\bf p})$ defined via\cite{4}
\begin{equation}
|B({\bf x})\rangle =  \int d^{3}p \varphi({\bf p}) e^{i{\bf p\cdot x}}
|B({\bf p})\rangle
\end{equation}
with normalization condition\footnote{For Dirac spinors we use the
convention $u({\bf p},s)=\sqrt{E+m} \left( \begin{array}{c} \chi_{s} \\
\frac{\sigma \cdot p}{E+m} \chi_{s} \end{array} \right)$ \cite{4} }
\begin{eqnarray}
\int d^{3}x \;\langle \bar{B} ({\bf x})|B({\bf x})\rangle
& = & \int d^{3}p \; 2M \; (2\pi)^{3} |\varphi({\bf p})|^{2} \nonumber \\
& = & 1 \label{eq:norm1}
\end{eqnarray}
Note that the hadron bag state $|B({\bf x})\rangle$ is centered about point
${\bf x}$ in position space. Center of mass motion of the bag due to the
internal
quark dynamics will not be taken into account.

Suppose that one is interested in the time component of the momentum space
matrix element. {\it i.e.}
\begin{eqnarray}
A_{0}^{WP} & = & \int d^{3}x d^{3}p' d^{3}p \varphi^{\bf *}({\bf p'})
                  \varphi ({\bf p}) J_{0}^{\it V}({\bf p},{\bf p'})
                  e^{iq\cdot x} \nonumber\\
            & = & f_{1}(0) \label{eq:time1}
\end{eqnarray}
In the quark model we can also calculate this transition moment,
yielding
\begin{equation}
A_{0}^{\it QM} = _{QM}\langle p\uparrow|\int d^{3}x \; \bar{\psi_{u}}({\bf
x}) \gamma_{0} \psi_{d}({\bf x}) |n \uparrow \rangle_{QM},
\label{eq:L0}
\end{equation}
where $|N \rangle_{QM}$ represent the quark model state vectors of the
nucleon while $\psi_{q}({\bf x})$ denote quark field operators. As
we consider only S-wave quark states throughout, we can represent the field
operators as
\begin{equation}
\psi_{q}(x)=\sum_{spin}[\phi_{0,s}({\bf x})e^{-i\omega_{0}t}b_{q}(s)
+\phi^{\dagger}_{0,s'}({\bf x})e^{i\omega_{0}t} b_{q}^{\dagger}(s')],
\end{equation}
where
\begin{equation}
\phi_{0,s}({\bf x}) = \left( \begin{array}{c} i u({\bf x}) \chi_{s} \\
l({\bf x}) {\bf \sigma \cdot \hat{x}} \chi_{s} \end{array} \right).
\label{eq:co}
\end{equation}
Here $u({\bf x}), l({\bf x})$ correspond to upper, lower components of
the quark wavefunctions respectively and are normalized via
\begin{equation}
\int d^{3}x \phi^{\dagger}({\bf x})\phi ({\bf x}) = \int d^{3}x
(u^{2}({\bf x})+l^{2}({\bf x})) = 1.
\end{equation}
Using these expressions we can evaluate Eq.~\ref{eq:L0}  yielding
\begin{eqnarray}
A_{0}^{\it QM} & = & \int d^{3}x \left( u_{u}({\bf x}) u_{d}({\bf x}) +
                      l_{u}({\bf x}) l_{d}({\bf x}) \right) \nonumber\\
                & = & 1,
\end{eqnarray}
in the limit of SU(2) symmetry. Lastly, equating quark model and wavepacket
expressions we find
\begin{equation}
f_{1}(0)=1,
\end{equation}
as required by CVC.

In order to evaluate the weak magnetic form factor---$f_{2}(0)$---we take a
first moment of the matrix element, yielding
\begin{eqnarray}
A_{1}^{WP} & = & \int d^{3}x d^{3}p' d^{3}p \varphi^{\bf *}({\bf p'})
                  \varphi ({\bf p})\frac{1}{2} \epsilon_{ij3} x^{i}
                  J^{j}_{\it V}({\bf p},{\bf p'}) e^{iq\cdot x}
                  \label{eq:wavepacket} \nonumber\\
            & = & \frac{f_{1}(0)+f_{2}(0)}{2M}, \label{eq:M1hadronic}
\end{eqnarray}
where $\epsilon_{ijk}$ is the completely antisymmetric Levi-Civita tensor.
Again by CVC we require\cite{3}
\begin{equation}
f_{1}(0) + f_{2}(0) = 1 + \kappa_{p} - \kappa_{n} = 4.7,
\end{equation}
where $\kappa_{i}$ is the anomalous magnetic moment of the nucleon $i$.

On the other hand, in the constituent quark model,
\begin{eqnarray}
A_{1}^{\it QM}  & = &  _{QM}\langle p\uparrow|\int d^{3}x \; \frac{1}{2}
                      \epsilon_{ij3} x^{i} \left[ \bar{\psi_{u}}(x) \gamma^{j}
                      \psi_{d}(x) \right] |n \uparrow \rangle_{QM} \nonumber\\
                & = & \frac{5}{9} \int d^{3}x \; |\vec{x}| [ u_u({\bf x})
                      l_d({\bf x}) + u_d({\bf x}) l_u({\bf x}) ].
\end{eqnarray}
Equating wavepacket and quark model results we have the prediction\footnote
{For example, using MIT bag model wavefunctions\cite{5} one finds
\begin{equation}
f_{2}(0)=2.5,
\end{equation}
which, when center of mass corrections are included, is boosted to\cite{6}
\begin{equation}
f_{2}(0)=3.2,
\end{equation}
and is in reasonable agreement with the CVC and experimental value\cite{7}
\begin{equation}
f_{2}^{\it CVC}(0)=\kappa_{p} -\kappa_{n} = 3.7.
\end{equation}
However, most of our considerations below will be independent of specific
choices for quark wavefunctions.}
\begin{equation}
f_{2}^{\it QM}= \frac{10M}{9} \int d^{3}x \; |\vec{x}| [ u_u({\bf x}) l_d({\bf
x})+u_d({\bf x})l_u({\bf x}) ]  - 1.  \label{eq:f2}
\end{equation}

Next we turn to the corresponding N-$\Delta$ polar vector transition, for
which the most general matrix element can be written, using spin-parity
considerations, in terms of four form factors $c_{i}(q^{2})$ as
\begin{eqnarray}
J^{\it V}_{\mu \; \Delta^{++} p}(\bf{p},\bf{p'})
& = & \langle \Delta^{++} ({\bf p'})|J^{\it V}_{\mu \; \Delta p}|p({\bf p})
      \rangle  \nonumber \\
& = & \bar{\Delta}^{\nu \; ++} ({\bf p'}) \mbox{\LARGE [(}
      \frac{c_{1}(q^{2})}{2M} \gamma^{\lambda} + \frac{c_{2}(q^{2})}{4M^{2}}
      q^{\lambda} + \frac{c_{3}(q^{2})}{4M^{2}} p^{\lambda}
      \mbox{\LARGE )} \nonumber \\
&   & \times ( q_{\lambda} g_{\mu \nu} - q_{\nu} g_{\lambda \mu} ) +
      c_{4}(q^{2}) g_{\mu \nu}  \mbox{\LARGE ]}\gamma_5 u({\bf p}).
      \label{eq:veccur2}
\end{eqnarray}
Here the $\Delta^{++}$ is represented in terms of a Rarita-Schwinger spinor
\cite{8} and the momentum transfer is defined as $q_{\mu}=p'_{\mu}
- p_{\mu}$. Using CVC we require $c_{4}(q^{2})=0$.
In order to extend the wavepacket formalism to delta states, represented as
free particle Rarita-Schwinger spinors, we define a wavepacket function
$\rho ({\bf p})$ via
\begin{equation}
|\Delta_{\mu} ({\bf x}) \rangle = \int d^{3}p \; \rho ({\bf p}) e^{i {\bf p
\cdot x}} |\Delta_{\mu} ({\bf p}) \rangle,
\end{equation}
normalised as
\begin{eqnarray}
\int d^{3}x \; \langle \bar{\Delta}^{\mu} ({\bf x})| \Delta_{\mu} ({\bf x})
\rangle & = & \int d^{3}p \; 2 M_{\Delta} \; (2\pi)^{3} |\rho ({\bf
              p})|^{2} \nonumber \\
        & = & 1. \label{eq:norm2}
\end{eqnarray}

Comparing Eq.~\ref{eq:norm1} and Eq.~\ref{eq:norm2} we make the following
identification between the two wavepacket functions for our calculations of
$N-\Delta$ transition moments:
\begin{equation}
\varphi ({\bf p}) \sqrt{2M} = \rho ({\bf p}) \sqrt{2M_{\Delta}}
\label{eq:wpmatch}
\end{equation}
Unlike in the case of $N-N$ transitions (Eq.~\ref{eq:time1}) the time
component of Eq.~\ref{eq:veccur2} vanishes in both wavepacket and quark model
calculations, as expected. Note that in our use of the wavepacket formalism
the three-momenta of initial and final particle are forced to be the same.
\footnote{If we want to account explicitly for the four-momentum $k_{\mu}$
transferred to the initial hadron bag by a photon or a $W^{\pm}$ the
situation gets much more complicated. For a generic matrix element $M$ and
a generic form factor $F(q^{2})$ we have
\begin{eqnarray}
M^{\it WP'} & = & \int d^{3}x d^{3}p d^{3}p' \rho^{*}({\bf p'}) \varphi({\bf
                      p}) F((p'-p)^{2}) e^{-i({\bf p'}-{\bf p}-{\bf k})
                      \cdot{\bf x}} \nonumber \\
                & = & (2\pi)^{3} \int d^{3}p \; \rho^{*}({\bf p}+{\bf k})
                      \; \varphi({\bf p}) \; F((E'-E)^{2}-{\bf k}^{2})
\end{eqnarray}
Note that in this case we cannot make use of Eq. 22 because
the two wavepacket functions $\varphi({\bf p})$ and $\rho({\bf p})$
are not evaluated at the same momentum. In order to solve this problem we
would have to model one of these functions explicitly. In our calculations
we therefore set ${\bf k}=0$ and approximate the static bag form factor
$F(q^{2}) \rightarrow F((E'-E)^{2}) \approx F((M'-M)^{2})$

One way to avoid this extra model dependence is to account for recoil of the
final state hadron bag. This was done for some of our calculations \cite{30}
and the effects were found to be of the same order as the static bag's center
of mass corrections. We therefore feel justified to neglect the factor
$\exp(i{\bf k}\cdot{\bf x})$ in our calculations.}  In the case of
nucleon-delta transitions this implies $q^{2}=q_{0}^{2}\neq 0$ due to their
differing mass.  For the first non-vanishing moment we find in the wavepacket
formalism
\begin{eqnarray}
B_{1}^{WP}  & = & \int d^{3}x d^{3}p' d^{3}p \; \rho^{\bf *}({\bf p'})
                  \varphi ({\bf p}) \; \frac{1}{2} \epsilon_{ij3} x^{i}
                  J_{\Delta^{++}p}^{j {\it V}}({\bf p},{\bf p'})
                  e^{iq\cdot x} \nonumber \\
            & = & \sqrt{1\over 6}\frac{M+M_{\Delta}}{2M_{\Delta}}
                  \mbox{\LARGE [} \frac{c_{1}(q_{0}^{2})}{2M}
                  \left( 1+ \frac{M_{\Delta} + M}{2M} \right) +
                  \frac{c_{2}(q_{0}^{2})}{2M} \frac{q_{0}^{2}}{4M^{2}}
                  \nonumber \\
            &   & + \frac{c_{3}(q_{0}^{2})}
                  {2M} \left( \frac{M_{\Delta}^{2}-M^{2}-q_{0}^{2}}{8M^{2}}
                  \right) \mbox{\LARGE ]} \label{eq:M1hadronic2}
\end{eqnarray}
with $q_{0}=M_{\Delta}-M$ and
\begin{eqnarray}
B_{1}^{\it QM}  & = & _{QM}\langle \Delta^{++} \uparrow|\int d^{3}x \;
                      \frac{1}{2} \; \epsilon_{ij3} x^{i} \left[ \bar{\psi_{u}}
                      (x) \gamma^{j} \psi_{d}(x) \right] |p
                      \uparrow \rangle_{QM} \nonumber\\
                & = & \frac{4}{3}\sqrt{1\over 6} \int d^{3}x |\vec{x}| [ u_d({
                      \bf x}) l_u({\bf x}) + u_u({\bf x})l_d({\bf x}) ]
\end{eqnarray}
in the quark model.

Equating these expressions we have the prediction
\begin{eqnarray}
c_{1}(q_{0}^{2}) \left( 1+ \frac{M_{\Delta} + M}{2M}
\right) + c_{2}(q_{0}^{2}) \frac{(M_{\Delta}-M)^{2}}{4M^{2}} + c_{3}(q_{0}^{2})
\frac{M_{\Delta}-M}{4M} \nonumber \\
= \frac{2M_{\Delta}}{M+M_{\Delta}} \times \frac{8M}{3} \int d^{3}x
|\vec{x}| [ u_d({\bf x}) l_u({\bf x})+u_u({\bf x})l_d({\bf x}) ].
\end{eqnarray}
However, rather than use specific quark wavefunctions, we can employ
Eq.~\ref{eq:f2} to write the equivalent form
\begin{eqnarray}
c_{1}(q_{0}^{2}) \left( 1+ \frac{M_{\Delta} + M}{2M}
\right) + c_{2}(q_{0}^{2}) \frac{(M_{\Delta}-M)^{2}}{4M^{2}} + c_{3}(q_{0}^{2})
\frac{M_{\Delta}-M}{4M} \nonumber \\
= \frac{2M_{\Delta}}{M+M_{\Delta}} \times  \frac{12}{5} \left( 1 + f_{2}(0).
\right) \label{eq:prediction}
\end{eqnarray}
We note that in the limit $M_{\Delta}=M$ this becomes the
familiar SU(6) prediction
\begin{equation}
c_{1}(0)=\frac{6}{5} ( 1+ f_{2}(0) ).
\end{equation}

We can proceed to calculate a second moment of Eq.~\ref{eq:veccur2}:
\begin{eqnarray}
B_{2}^{WP}  & = & \int d^{3}x d^{3}p' d^{3}p \; \rho^{\bf *}({\bf p'})
                  \varphi ({\bf p}) \; [ 3 x_{3}^{2} - \vec{x}^{2} ]
                  J_{\Delta^{++}p}^{0 {\it V}}({\bf p},{\bf p'})
                  e^{iq\cdot x} \nonumber \\
            & = & \sqrt{\frac{3}{2}} \frac{(M+M_{\Delta})}{2M_{\Delta}}
                  \frac{1}{M^{2}} \left[ c_{1}(q_{0}^{2}) + \frac{M_{\Delta}-M}
                  {2M} c_{2}(q_{0}^{2}) + \frac{1}{2} c_{3}(q_{0}^{2}) \right].
                  \label{eq:E2}
\end{eqnarray}
However, in the constituent quark model, this moment vanishes as
our quark wavefunctions (Eq.~\ref{eq:co}) are purely S-wave.
\begin{eqnarray}
B_{2}^{\it QM}  & = & _{QM}\langle \Delta^{++} \uparrow|\int d^{3}x \;
                      [ 3 x_{3}^{2} - \vec{x}^{2} ] \bar{\psi_{u}}
                      (x) \gamma^{0} \psi_{d}(x) |p
                      \uparrow \rangle_{QM} \nonumber\\
                & = & 0. \label{eq:E2quark}
\end{eqnarray}
Experimentally this restriction turns out to be well justified because
the vector part of the nucleon-delta transition is dominated by the M1
amplitude \cite{9}. Combining the results of Eq.~\ref{eq:E2},
Eq.~\ref{eq:E2quark} and Eq.~\ref{eq:prediction} we find the prediction
\begin{equation}
c_{1}(q_{0}^{2})= \frac{2M_{\Delta}}{M+M_{\Delta}} \times  \frac{6}{5} \left(
1 + f_{2}(0) \right).
\end{equation}
In order to have everything written in terms of experimentally accessible
quantities the only thing left to do is the scaling down of the form factor
$c_{1}(q^{2})$ from the time-like point
$q_{0}^{2}=(M_{\Delta}-M)^{2}=0.086$GeV$^{2}$ to the photon-point $q^{2}=0$.
For the $q^{2}$ behavior we use the empirical parametrization\cite{9a}
\begin{equation}
c_{1}(q^{2}) = \frac{1}{\sqrt{1-\frac{q^{2}}{1.43\mbox{\tiny GeV}^{2}}}}
               \times \frac{c_{1}(0)}{(1-\frac{q^{2}}{0.71\mbox{\tiny GeV}^{
               2}})^{2}},
\end{equation}
which, when extrapolated to the time-like region, yields
\begin{eqnarray}
c_{1}(0)^{\it th.} & = & \frac{1}{1.34} \; \frac{2M_{\Delta}}{M+M_{\Delta}}
                         \times  \frac{6}{5} \left( 1 + f_{2}(0) \right)
                         \nonumber \\
                   &\approx& 4.8. \label{eq:c1}
\end{eqnarray}

In order to see how well these predictions work, we note that $c_{1}(0),
c_{3}(0)$ can be determined by use of CVC and the analogous electromagnetic
transition $\gamma p \rightarrow \Delta^{+}$ for which the most general
gauge-invariant matrix element has the form
\begin{eqnarray}
M_{\it \gamma p \Delta^{+}} = -e \sqrt{\frac{2}{3}} \Delta^{+}_{\mu}
({\bf p_{\Delta}}) \mbox{\LARGE [} \frac{h_{1}(0)} {2M} \mbox{\LARGE (}
(M_{\Delta} + M) \epsilon^{\mu} + \gamma \cdot \epsilon \; p_{N}^{\mu}
\mbox{\LARGE )} \nonumber \\
+ \frac{h_{3}(0)}{4M^{2}} \mbox{\LARGE (}
\frac{1}{2} (M_{\Delta}^{2} - M^{2}) \epsilon^{\mu} + p_{N}
\cdot \epsilon \; p_{N}^{\mu} \mbox{\LARGE )]} \gamma_{5} u({\bf p_{N}})
\end{eqnarray}
where $\epsilon^{\mu}$ denotes the polarization four-vector of the (real)
photon.
{}From photoproduction experiments in the $\Delta$ region one
extracts\cite{9}
\begin{equation}
h_{1}^{\it exp.}(0) = 5.10 \pm 0.55, \quad h_{3}^{\it exp.}(0)
= - 5.41 \pm 0.85.
\end{equation}
Then, using the CVC constraint
\begin{equation}
c_{i}(0) = \left[ \sqrt{\frac{2}{3}} h_{i}(0) \right] \cdot \sqrt{3},
\end{equation}
one finds
\begin{eqnarray}
c_{1}(0)^{\it exp.} =& &7.21 \pm 0.78, \\
c_{3}(0)^{\it exp.} =&-&7.65 \pm 1.20.
\end{eqnarray}
We note that the result for $c_{1}(0)$ is more than 30\% larger than its quark
model mass-corrected SU(6) prediction Eq.~\ref{eq:c1}. This result is rather
surprising given the relatively simple physics
involved and traditional gluonic hyperfine effects which explain the the
$N,\Delta$ splitting are unable to account for this excess M1 strength in the
N-$\Delta$ region \cite{10}, although
we shall return to this point later.

\subsection{The Axial Form Factors}

Moving to the axial vector current, we define the general axial matrix
element between neutron and proton in terms of three structure functions
$g_{i}(q^{2})$:
\begin{eqnarray}
J_{\mu}^{\it A}({\bf p},{\bf p'}) & = & \langle p({\bf p'})|A_{\mu}|n({\bf p})
                                        \rangle \nonumber \\
                                  & = & \bar{u}({\bf p'})\left[ g_{1}(q^{2})
                                        \gamma_{\mu} + \frac{i g_{2} (q^{2})}
                                        {2M} \sigma_{\mu \nu} q^{\nu} +
                                        \frac{g_{3}(q^{2})}{2M} q_{\mu}
                                        \right] \gamma_{5} u({\bf p}).
                                        \label{eq:axcur1}
\end{eqnarray}
In the SU(2) limit, $g_{2}(q^{2})=0$ from the
G-invariance considerations\cite{13}.
Also $g_{3}(q^{2})$ contains the pion pole, and is, strictly speaking, outside
the simple constituent quark model \cite{14}. Finally, in the case of
$g_{1}(q^{2})$
we find
\begin{eqnarray}
C_{0}^{\it WP} & = & \int d^{3}x d^{3}p d^{3}p' \varphi^*({\bf p}')
                     \varphi{\bf p}) J^{3}_{\it A}({\bf p},
                     {\bf p'})e^{i q\cdot x} \nonumber \\
               & = & g_{1}(0), \label{eq:M0a}
\end{eqnarray}
in the wavepacket approach, and
\begin{eqnarray}
C_{0}^{\it QM} & = & _{QM}\langle p \uparrow|\int d^{3}x \; \bar{\psi}_{u}(x)
                     \gamma^{3} \gamma_{5} \psi_{d}(x)|n \uparrow
                     \rangle_{QM} \nonumber \\
               & = & \frac{5}{3} \left[ 1 - \frac{4}{3} \int d^{3}x \;
                     l_u({\bf x}) l_d({\bf x}) \right],
                     \label{eq:M0quark}
\end{eqnarray}
in the constituent quark model. Equating these expressions, we find
\begin{equation}
g_{1}(0)=\frac{5}{3} \left[ 1 - \frac{4}{3} \int d^{3}x \;
l_u({\bf x}) l_d({\bf x}) \right]. \label{eq:np}
\end{equation}
If we set $l_i({\bf x})=0$, we recover the well-known SU(6) result
$g_{1}(0)=\frac{5}{3}$, in the nonrelativistic quark
model\cite{15}. It is the inclusion of the lower component
of the relativistic wavefunctions which brings this value down to the
experimental number $g_{1}^{\it exp.}(0)=1.262 \pm 0.004$\cite{16}. Thus
we require\footnote {We note that this result is approximately
obtained by
using the MIT bag wavefunctions\cite{6}.
However, we do not wish to specify any specific model.}
\begin{equation}
\int d^{3}x \; l_u({\bf x}) l_d({\bf x}) \approx 0.4.
\end{equation}

For the axial N-$\Delta$ transition, the matrix element can be written from
spin-parity arguments in terms of four form factors $d_{i}(q^{2})$
\begin{eqnarray}
J_{\mu \Delta N}^{\it A}({\bf p},{\bf p'})
& = & \langle \Delta^{++} ({\bf p'})|A_{\mu}^{\Delta}|p({\bf p}) \rangle
       \nonumber \\
& = & \bar{\Delta}^{++ \nu} ({\bf p'}) \mbox{\LARGE [} d_{1}(q^{2}) g_{\mu
      \nu} + \frac{d_{2}(q^{2})}{M^{2}} P^{\alpha} \left( q_{\alpha} g_{\mu
      \nu} - q_{\nu} g_{\alpha \mu} \right) \nonumber \\
&   & \hspace{13 mm} - \frac{d_{3}(q^{2})}{M^{2}} p_{\nu} q_{\mu} + i
      \frac{d_{4}(q^{2})}{M^{2}} \epsilon_{\mu \nu \alpha \beta} P^{\alpha}
      q^{\beta} \gamma_{5} \mbox{\LARGE ]} u({\bf p}), \label{eq:axcur2}
\end{eqnarray}
where  $P_{\mu}=p'_{\mu} + p_{\mu}$ and $q_{\mu}=p'_{\mu}- p_{\mu}$. (In
Appendix A, we give the connection between the form factors $d_{i}(q^{2})$ and
the $C_{i}^{\it A}(q^{2})$ often used in previous works\cite{17}.)

Equating the wavepacket result
\begin{eqnarray}
D_{0}^{\it WP} & = & \int d^{3}x d^{3}p d^{3}p' \rho^*
                            ({\bf p}')\varphi({\bf p})\; J_{
                                 \Delta N}^{3 {\it A}}({\bf p},{\bf p'})
                                 e^{i q\cdot x} \nonumber \\
                           & = & \sqrt{\frac{2}{3}} \left( d_{1}(q_{0}^{2}) +
                                 d_{2}(q_{0}^{2}) \frac{M_{\Delta}^{2} - M^{2}}
                                 {M^{2}} \right),
\end{eqnarray}
where  $q_{0}=M_{\Delta}-M$, with that given in the corresponding quark model
calculation
\begin{eqnarray}
D_{0}^{\it QM} & = & _{QM}\langle \Delta^{++} \uparrow|\int d^{3}x
                                \; \bar{\psi}_{u}(x) \gamma^{3} \gamma_{5}
                                \psi_{d}(x)|p \uparrow \rangle_{QM}
                                \nonumber \\
                          & = & 2 \sqrt{\frac{2}{3}} \left[ 1 - \frac{4}{3}
                                \int d^{3}x \; l_u({\bf x}) l_d({\bf x})
                                \right],
\end{eqnarray}
we find
\begin{equation}
d_{1}(q_{0}^{2}) + d_{2}(q_{0}^{2}) \frac{M_{\Delta}^{2} - M^{2}}{M^{2}} = 2
\left[ 1-\frac{4}{
3} \int d^{3} x l_{u}({\bf x}) l_{d}({\bf x}) \right]. \label{eq:d1}
\end{equation}
As in the case of the vector formfactors, we want to scale the
$d_{i}(q^{2})$ from $q_{0}^{2}=(M_{\Delta}-M)^{2}=0.086$GeV$^{2}$ down to
the photon-point $q^{2}=0$. For the $q^{2}$ dependence we use the
empirical parametrisation\cite{22}
\begin{equation}
d_{i}(q^{2})=d_{i}(0)\frac{1+1.21\frac{q^{2}}{2\mbox{\tiny GeV}^{2} -
q^{2}}}{(1-\frac{q^{2}}{M_{A}^{2}})^{2}},
\end{equation}
where $M_{A}$ ranges from $1.14$ to $1.28$ GeV, and extrapolate it
to the timelike region:
\begin{equation}
d_{1,2}(q_{0}^{2})\approx d_{1,2}(0) \times (1.17 \pm 0.03).
\end{equation}

Comparing with the corresponding expression for the neutron-proton
transition (Eq.~\ref{eq:np}) we can now write Eq.~\ref{eq:d1} entirely in terms
of experimental quantities as
\begin{equation}
d_{1}(0) + d_{2}(0) \; \frac{M_{\Delta}^{2} - M^{2}}{M^{2}} =
\frac{d_{1,2}(0)}{d_{1,2}(q_{0}^{2})}\times \frac{6}{5} \; g_{1}(0),
\label{eq:axial}
\end{equation}
which can be tested in charged current neutrino-nucleon scattering. In the
degenerate limit $M_{\Delta}=M$, this becomes the well-known SU(6)
relation $d_{1}(0)=\frac{6}{5} \; g_{1}(0)$.

Before
confronting this prediction with experiment, we examine additional relations
which arise in the quark model.  As in the nucleon case
$d_{3}(q^{2})$ contains the pion pole and is outside the simple
constituent quark formalism.  In
the case of $d_{4}(q^{2})$,
 our use of the S-wave wavefunctions yields $d_{4}
(q^{2})=0$, which is consistent with other calculations ({\it cf.} appendix A).
Finally, calculating a first moment of the axial current, and omitting
$d_{3}(q^{2})$ contributions, we find
\begin{eqnarray}
D_{1}^{\it WP} & = & - i \int d^{3}x d^{3}p d^{3}p' \; x_{3} \rho^*({\bf p}')
                     \varphi({\bf p})\;J_{\Delta N}^{0 {\it A}}({\bf p},{\bf
                     p'})e^{i q\cdot x} \nonumber \\
               & = & \sqrt{\frac{2}{3}} \left( \frac{d_{1}(q_{0}^{2})}{2M_{
                     \Delta}} + d_{2}(q_{0}^{2}) \frac{M+M_{\Delta}}
                     {M^{2}} \right). \label{eq:E1}
\end{eqnarray}
The corresponding quark model calculation of this moment
gives
\begin{eqnarray}
D_{1}^{\it QM}           & = & - i \; _{QM}\langle \Delta^{++} \uparrow|\int
                                d^{3}x \; x_{3} \; \bar{\psi}_{u}(x)
                                \gamma_{0} \gamma_{5} \psi_{d}(x)|p \uparrow
                                \rangle_{QM} \nonumber \\
                          & = & \frac{2}{3} \sqrt{\frac{2}{3}} \int d^{3}x
                                \; x \left[ u_{u}({\bf x}) l_{d}({\bf x}) -
                                u_{d}({\bf x}) l_{u}({\bf x}) \right]
                                \nonumber \\
                          &\approx& 0, \label{eq:E1quark}
\end{eqnarray}
which is a consequence of SU(2) symmetry\cite{18} and brings about an
additional relation between $d_{1}(0)$ and $d_{2}(0)$:
\begin{equation}
d_{1}(0) + 2 \; d_{2}(0) \; \frac{M_{\Delta} (M+M_{\Delta})}{M^{2}} = 0.
\label{eq:edm}
\end{equation}
Using this result Eq.~\ref{eq:axial} can be written in the simpler form
\begin{eqnarray}
d_{1}(0)^{\it th.} & = & \frac{1}{1.17} \times \frac{6}{5} \; g_{1}(0) \;
                         \frac{2M_{\Delta}}{M_{\Delta} +
M},\label{eq:quarkd1}\\
d_{2}(0)^{\it th.} & = & - \frac{1}{1.17} \times \frac{6}{5} \; g_{1}(0) \;
                         \frac{M^{2}}{(M+M_{\Delta})^{2}}. \label{eq:quarkd2}
\end{eqnarray}

We now examine the experimental determination of the
axial N-$\Delta$ transition form factors. In previous papers it was assumed
that $d_{1}(0)$ is the only major contributing
form factor to the cross section $\nu p \rightarrow \Delta^{++} \mu^{-}$,
at low $q^{2}$.
For example, Barish {\it et al.} find\cite{19}
\begin{equation}
d_{1}^{\it exp.}(0)=2.0 \pm 0.4. \label{eq:d1barish}
\end{equation}
We show their low $q^{2}$ data points in Figure 1. Note the large
uncertainty in this result but also that it is consistent with the old model
calculation by Adler\cite{20}, as newly parametrized by Schreiner
{\it et al.}\cite{21}.

In Figure 2 we show the results of a more recent experiment by Kitagaki et
al.\cite{22} using a neutrino beam of $<E_{\nu}> \approx 1.6$ GeV and a
deuterium target. They measured the ratio of two exclusive
neutrino cross sections and
found
\begin{equation}
R\equiv \frac{\frac{d\sigma}{dq^{2}} (\nu d \rightarrow \mu^{-} \Delta^{++}
n)|_{q^{2}
\approx 0}}{\frac{d\sigma}{dq^{2}} (\nu d \rightarrow \mu^{-} p p)|_{q^{2}
\approx 0}} = 0.50 \pm 0.05, \label{eq:Kitagaki}
\end{equation}
for the point of lowest $q^{2}=0.1 \mbox{ GeV}^{2}$. Stimulated by this result,
we note that the corresponding cross sections can be expressed, at $q^2=0,$ as
\begin{eqnarray}
\frac{d\sigma}{dq^{2}}(\nu n \rightarrow \mu^{-} p)|_{q^{2}=0}
& = & \cos^{2}\theta_{C} \frac{G_{F}^{2}}{2\pi} \left( f_{1}^{2}(0) +
      g_{1}^{2}(0) \right), \nonumber\\
&   & \nonumber \\
\frac{d\sigma}{dq^{2}}(\nu p \rightarrow \mu^{-} \Delta^{++})|_{q^{2}=0}
& = & \cos^{2}\theta_{C} \frac{G_{F}^{2}}{6\pi} \;
      \frac{(M_{\Delta}+M)^{2}}{M_{\Delta}^{2}+M^{2}} \; \frac{2ME_{\nu}+(M^{2}
      -M_{\Delta}^{2})}{2ME_{\nu}} \nonumber \\
&   & \hspace{15 mm} \cdot \frac{M^{2}}{M_{\Delta}^{2}} \; \left[ d_{1}(0) +
      d_{2}(0) \frac{M_{\Delta}^{2} - M^{2}}{M^{2}} \right]^{2},
\end{eqnarray}
where $G_{F}$ denotes the Fermi constant and $\cos\theta_{C}$ represents
$V_{ud}$ in the KM-matrix. (Note that we have here neglected all terms
depending on the mass of the muon.) Extrapolating the experimental result of
Eq.~\ref{eq:Kitagaki} to $q^{2}=0$, we find
\begin{eqnarray}
d_{1}(0) + d_{2}(0) \frac{M_{\Delta}^{2} - M^{2}}{M^{2}}
& = & \sqrt{R} \; \sqrt{6} \; \sqrt{f_{1}^{2}(0) + g_{1}^{2}(0)} \; \frac{
      \sqrt{M_{\Delta}^{2}+M^{2}}}{M_{\Delta}+M}  \nonumber \\
&   & \hspace{15 mm} \times \frac{M_{\Delta}}{M} \; \sqrt{\frac{M E_{\nu}}{2 M
      E_{\nu} + (M^{2} - M_{\Delta}^{2})}}.
\end{eqnarray}
The two main uncertainties in this result come from the spread in the neutrino
beam energy and from the experimental error
bar in Eq.~\ref{eq:Kitagaki}. Depending on whether we use the mean neutrino
beam energy $<E_{\nu}>=1.6$ GeV or the peak in the neutrino beam energy
distribution $E_{\nu}^{\rm pk}=1.2$ GeV, one gets
\begin{eqnarray}
d_{1}(0) + d_{2}(0) \frac{M_{\Delta}^{2} - M^{2}}{M^{2}}
& = & 2.08 \pm 0.10  ( E_{\nu}=1.6 \mbox{ GeV} ), \nonumber\\
& = & 2.18 \pm 0.11  ( E_{\nu}=1.2 \mbox{ GeV} ).
\end{eqnarray}
Note that the error bars include only those of Eq.~\ref{eq:Kitagaki}.
One can see that the uncertainty coming from the neutrino beam energy is of
the same magnitude. We, therefore, conclude
\begin{equation}
d_{1}(0) + d_{2}(0) \frac{M_{\Delta}^{2} - M^{2}}{M^{2}} = 2.1 \pm 0.2.
\end{equation}
Use of Eq.~\ref{eq:edm} enables us to determine $d_{1}(0)$ and $d_{2}(0)$
independently:
\begin{eqnarray}
d_{1}^{\it exp.}(0) & = & 2.1 \; \frac{2M_{\Delta}}{M_{\Delta}+M} \nonumber \\
                    & = & 2.4 \pm 0.25, \label{eq:d1kitagaki} \nonumber\\
                    &   & \nonumber \\
d_{2}^{\it exp.}(0) & = & - 2.1 \; \frac{M^{2}}{(M+M_{\Delta})^{2}}
                          \nonumber \\
                    & = & - 0.4 \pm 0.05.
\end{eqnarray}
Comparing these experimental results with the predictions of the quark
model (Eq.~\ref{eq:quarkd1}, Eq.~\ref{eq:quarkd2})
\begin{equation}
d_{1}^{\it QM}(0)=1.5 \pm 0.05 \quad\mbox{and} \quad d_{2}^{\it QM}(0)=-0.25
\pm
0.005,
\end{equation}
we come to the conclusion that once more
the quark model significantly underestimates the
strength of the axial transition form factors---for the dominant form factor
$d_{1}(0)$, we find a deviation of more than 35\% !  (Note that the result of
Kitagaki {\it et al.} (Eq.~\ref{eq:d1kitagaki}) is consistent with
the previous
determination by Barish {\it et al.} (Eq.~\ref{eq:d1barish}), though the
central
values are somewhat different.)

Having now examined the axial vector amplitudes for NN and N$\Delta$
transitions, we move on to the analogous strong pion couplings.

\subsection{Strong Coupling Constants}

In the case of the strong couplings, we begin by defining effective
Lagrangians for the $\pi NN$ and $\pi N\Delta$ interaction
\begin{eqnarray}
     {\cal L}_{\it \pi NN} & = & -i g_{\pi NN} \; \bar{u} ({\bf p'}) \gamma_{5}
                                 {\bf \tau}^{i} u({\bf p}) {\bf \pi}^{i},
                                 \nonumber\\
{\cal L}_{\it \pi N\Delta} & = & \frac{g_{\pi N\Delta}}{2M}
                                 \bar{\Delta}_{\mu}^{i} ({\bf p'}) g^{\mu \nu}
                                 u({\bf p}) \partial_{\nu} \pi^{i} + H.c..
                                 \label{eq:lag2}
\end{eqnarray}
Since the pion is a pseudoscalar, these both represent P-wave amplitudes,
which can be projected out via
\begin{equation}
M_{\it P-wave}^{\it WP}=\int d^{3}x d^{3}p' d^{3}p \; \varphi^{\ast}({\bf p'})
\varphi ({\bf p}) \; x_{3} \; M_{B'B\pi} ({\bf p},{\bf p'}) \; e^{i q \cdot x}.
\end{equation}
For the corresponding quark model multipole we find
\begin{equation}
M_{\it P-wave}^{\it QM} = - i N_{\pi} \; \int d^{3}x \; x_{3} \; _{QM}\langle
B'|\bar{\psi}_{u}(x)\gamma_{5} \psi_{d}(x)|B\rangle_{QM},
\end{equation}
where $N_{\pi}$ is an unknown normalization constant associated with the
created pion. Equating hadronic and quark model amplitudes, we find
\begin{eqnarray}
     g_{\pi NN}^{\it QM}(0) & = & \frac{10}{9 \sqrt{2}} M \; N_{\pi} \; \int
                               d^{3}x \; |\vec{x}| \left[ u_{u}(x) l_{d}(x) +
                   u_{d}(x) l_{u}(x) \right], \label{eq:strong1} \nonumber\\
g_{\pi N\Delta}^{\it QM}(q_{0}^{2}) & = & \frac{8}{3} \frac{M M_{\Delta}}{M+M_{
                \Delta}} \; N_{\pi} \; \int d^{3}x \; |\vec{x}| \left[ u_{u}(x)
                               l_{d}(x) + u_{d}(x) l_{u}(x) \right],
                               \label{eq:strong2}
\end{eqnarray}
with $q_{0}=M_{\Delta}-M$. In the limit $M_{\Delta}=M$ these relations reduce
to the familiar SU(6) prediction
\begin{eqnarray}
\frac{g_{\pi N\Delta}(0)}{g_{\pi NN}(0)} &\approx& \frac{g_{\pi
N\Delta}(m_{\pi}^{2})}{g_{\pi NN}(m_{\pi}^{2})} \nonumber \\
          &= & \frac{6\sqrt{2}}{5} = 1.70.
\end{eqnarray}
Assuming that the $q^{2}$ behavior of $g_{\pi N\Delta}$ scales like the
axial nucleon-delta transition form factor,
as given by the associated generalized
Goldberger-Treiman relation, we obtain the mass-corrected theoretical
prediction
\begin{eqnarray}
\frac{g_{\pi N\Delta}(0)}{g_{\pi NN}(0)} &\approx& \frac{g_{\pi
N\Delta}(m_{\pi}^{2})}{g_{\pi NN}(m_{\pi}^{2})} \nonumber \\
   &\approx& \frac{1}{1.17} \times \frac{6\sqrt{2}}{5} \frac{2M_{\Delta}}
       {M+M_{\Delta}} \nonumber \\
   &\approx& 1.6. \label{eq:strongratio}
\end{eqnarray}

Using the most recent value for the $\pi NN$ coupling\cite{11}
\begin{equation}
g_{\pi NN}^{\it exp.}=13.05 \pm 0.31,
\end{equation}
and the value of $g_{\pi N\Delta}$ extracted from a K-matrix analysis of
phase shifts\cite{9,12}
\begin{equation}
g_{\pi N\Delta}^{\it exp.}=28.6 \pm 0.3,
\end{equation}
we find
\begin{equation}
\frac{g_{\pi N\Delta}^{\it exp.}}{g_{\pi NN}^{\it exp.}} = 2.21 \pm 0.08,
\end{equation}
which again exceeds by 25\% the mass-corrected SU(6) prediction.

We now move on to discuss the Goldberger-Treiman relations which connect the
axial formfactors to the strong coupling constants.

\section{Goldberger-Treiman Relations}

Despite our inability to make direct contact between the strong
interactions of elementary particles and the QCD Lagrangian which
presumably underlies such interactions, it is possible to
exploit, at low energies, the (approximate) chiral symmetry
of QCD in order to provide
rigorous predictive power. In the NN sector, an example of this is the
Goldberger-Treiman relation\cite{2}
\begin{equation}
F_\pi g_{\pi NN}(q^2) =M g_{1}(q^{2}), \label{eq:gotr1}
\end{equation}
which is derived most simply via the PCAC relation
\begin{equation}
\partial^{\mu} A_{\mu}^{i} = F_{\pi} m_{\pi}^{2} \phi_{\pi}^{i},
\label{eq:pcac}
\end{equation}
assuming pion pole dominance of the pseudoscalar formfactor $g_{3}(q^{2})$,
which is consistent with a recent experiment\cite{23}.  Here $F_{\pi}=92.4$
MeV is the pion decay constant.  Note that the relation is strictly
valid only at the
{\it same} value of momentum transfer for both strong and axial couplings.
However,
in checking its experimental validity one generally uses $g_{1}(0)$ and
$g_{\pi NN}(m_{\pi}^{2})$. We thus expect a slight violation of GT, various
methods of taking this effect into account were analysed by
Dominguez\cite{24}. Defining
\begin{equation}
\Delta_{\pi} = 1 - \frac{M g_{1}(0)}{F_{\pi} g_{\pi NN}(m_\pi^2)},
\end{equation}
we anticipate $\Delta_{\pi} \approx 0.02$ from diagrams such as those in
Figure 3. Experimentally
things are not as clear, however, because of the presently uncertain value
of $g_{\pi NN}$ at $q^{2}=m_{\pi}^{2}$. The situation is summarized in Table 1,
and we see that things work to better than 5\%\cite{25} in any case .

\vspace{5 mm}
Table 1: \hspace{5 mm}
\begin{tabular}{|l|l|c|} \hline
                 &$g_{\pi NN}^{2}/4\pi$     & $\Delta_{\pi}$ \\ \hline
$\pi^{\pm}$      &$ 13.54\pm 0.05$ \cite{11}& 0.017          \\
                 &$ 13.31\pm 0.27$ \cite{26}& 0.008          \\
                 &$ 14.28\pm 0.18$ \cite{27}& 0.043          \\ \hline
$\pi^{0}$        &$ 13.47\pm 0.11$ \cite{11}& 0.014          \\
                 &$ 13.55\pm 0.13$ \cite{28}& 0.017          \\
                 &$ 14.52\pm 0.40$ \cite{29}& 0.051          \\ \hline
\end{tabular}

\vspace{5 mm}
Similarly one can derive the corresponding Goldberger-Treiman relation in
the N$\Delta$ sector\cite{30}. Using PCAC and assuming pion pole dominance
of the pseudoscalar form factor $d_{3}(q^{2})$ one finds
\begin{equation}
F_\pi g_{\pi N\Delta}(q^2)=\sqrt{2} M d_{1}(q^{2}).
\label{eq:gotr2}
\end{equation}
Again checking the validity of this result using $g_{\pi N\Delta}
(m_{\pi}^{2})$ and $d_{1}(0)$ one expects a violation of size
\begin{eqnarray}
\Delta_\pi^\Delta &=& 1 - {\sqrt{2} M d_{1}(0)\over F_{\pi}
g_{\pi N\Delta}(m_\pi^2)} \nonumber\\
&\approx & 0.02
\end{eqnarray}
where the size of $\Delta_{\pi}^{\Delta}$ is estimated using the diagrams
shown in Figure 4. Dillig and Brack came to similar results\cite{31}. Given
the large uncertainty in $g_{\pi N\Delta}$ we need not worry about this small
deviation at this point.

Using the value $d_{1}(0)$ extracted from neutrino scattering experiments
(Eq.~\ref{eq:d1barish} and Eq.~\ref{eq:d1kitagaki}), we can now try to test the
validity of the generalised Goldberger-Treiman relation Eq. 77 in
Table 2:
\vspace{5 mm}

Table 2: \hspace{10 mm}
\begin{tabular}{|l|c|} \hline
$d_{1}(0)$                       & "predicted" $g_{\pi N\Delta}$
 \\ \hline
$2.0\pm0.40$(Eq.~\ref{eq:d1barish})  &$29\pm 6$\\
\hline
$2.4\pm0.25$(Eq.~\ref{eq:d1kitagaki})&$34.5\pm 3.6$\\
\hline
\end{tabular}

\vspace{5 mm}
We observe that our extracted $d_{1}(0)$ (Eq.~\ref{eq:d1kitagaki}) indicates a
number at the upper end of
the present range for $g_{\pi N\Delta} \approx 26-33 $, but is consistent
with presently known information. We conclude that despite
the large violations of the mass-corrected SU(6) predictions for $d_{1}(0)$ and
$g_{\pi N\Delta}(0)$ the generalized Goldberger-Treiman relation
(Eq.~\ref{eq:gotr2}), demanded by chiral invariance, remains valid, within
errors. Since (broken) chiral symmetry is a property of the fundamental QCD
Lagrangian this result is perhaps not unexpected, but
is nonetheless reassuring.

\section{Conclusion}

In the previous sections we have examined the corrections which exist
between vector, axial and strong couplings between nucleons and
their counterparts in the N$\Delta$ sector. In the limit of degenerate
nucleon and delta masses, these relations are simply the result of SU(6)
symmetry. However, use of the constituent quark model provides significant
mass-dependent corrections to these predictions.
In each case the experimental N$\Delta$ coupling was found to be significantly
larger than its predicted value, by amounts ranging from 25\% to 35\%.
Despite these violations of the (mass-corrected) SU(6) predictions, the
connection between the axial and strong N$\Delta$ couplings required by
chiral symmetry---the Goldberger-Treiman relation---remains valid
in both NN and $N\Delta$ sectors.

A question which  has not been satisfactorily answered in
previous investigations is the origin of these surprisingly large symmetry
violations.  Conventional attempts to understand such SU(6) violations
via hyperfine gluonic interactions
have not been successful \cite{10}. However, the chiral solitton (Skyrmion)
model also yields model-independent relations
between diagonal and off-diagonal electroweak
form factors, and those for the strong couplings\cite{99}. These predictions
work much better in experimental tests, suggesting the important role
of degrees of freedom beyond those in the constituent quark model. It
is interesting to note that the tree-level effective Lagrangian in the
chiral soliton model already contains higher order physics derivable
from chiral perturbation theory \cite{99a}.

  This brings us to another promising line of thinking,
provided by heavy baryon techniques \cite{32,34}, wherein one
undertakes a rigorous expansion of transition amplitudes in terms of
powers of q/M.  Within this approach, one would expect the lowest order
parameters to obey the symmetries of the constituent quark model.  However,
in higher order, these quantities are renormalized by
the meson loop corrections
\begin{equation}
{\cal Q}\rightarrow{\cal Q}(1-\lambda {m_K^2\over 16\pi^2F_K^2}\ln ({m_K^2
\over \Lambda_\chi^2})),
\end{equation}
where $\lambda$ is a constant of ${\cal O}(1)$, which depends upon the process
being considered and $\Lambda_\chi\sim 1 GeV$ is the chiral scale parameter.
Since $m_K^2/16\pi^2F_K^2\sim 25\%$, such corrections have the potential to
lead to symmetry violations of the size found above.  However, as yet this
is only speculation and further work is needed in order to
test this hypothesis, as
will be reported in a future communication.

One of us (NCM) is grateful to L. Zhang for many useful discussions.

\appendix
\section{Notation}

Many experimental papers on neutrino induced delta production use
the notation of Llewellyn-Smith \cite{17} and Schreiner {\it et al.}\cite{21}.
In this notation the N-$\Delta$ vector transition current reads
\begin{eqnarray}
J_{\mu \Delta^{i} N}^{\it V}({\bf p},{\bf p'})
= \sqrt{3} \; \bar{\Delta}_{i}^{\nu}({\bf p'}) \mbox{\LARGE \{[} \frac{C_{
      3}^{\it V}(q^{2})}{M} \gamma^{\lambda} + \frac{C_{4}^{\it V}(q^{2})}
      {M^{2}} p'^{\lambda} + \frac{C_{5}^{\it V}(q^{2})}{M^{2}} p^{\lambda}
      \mbox{\LARGE ]} \nonumber \\
\hspace{40 mm} \times \left( q_{\lambda} g_{\mu \nu} - q_{\nu}
      g_{\lambda \mu} \right) + C_{6}^{\it V}(q^{2}) g_{\mu \nu}
      \mbox{\LARGE \}} \gamma_5u({\bf p}).
\end{eqnarray}
$M$ denotes the mass of the nucleon and the four momentum transfer is
defined as $q_{\mu}=p'_{\mu}-p_{\mu}$. Comparing with Eq.~\ref{eq:veccur2} we
find the following connections to our formfactors $c_{i}(q^{2})$:
\begin{eqnarray}
    c_{1}(0) & = & 2 \sqrt{3} C_{3}^{\it V}(0), \\
    c_{2}(0) & = & 4 \sqrt{3} C_{4}^{\it V}(0), \\
    c_{3}(0) & = & 4 \sqrt{3} ( C_{4}^{\it V}(0) + C_{5}^{\it V}(0) ), \\
c_{4}(q^{2}) & = & \sqrt{3} C_{6}^{\it V}(q^{2}) = 0. \label{eq:CVC}
\end{eqnarray}
Eq.~\ref{eq:CVC} results from CVC requirements.

For the N-$\Delta$ axial transition current experimentalists tend to use
\begin{eqnarray}
J_{\mu \Delta^{i}N}^{\it A}({\bf p},{\bf p'})
& = & \sqrt{3} \; \bar{\Delta}_{i}^{\nu}({\bf p'}) \mbox{\LARGE \{[}
      \frac{C_{3}^{\it A}(q^{2})}{M} \gamma^{\lambda} + \frac{C_{4}^{\it A}
      (q^{2})}{M^{2}} p'^{\lambda} \mbox{\LARGE ]} (q_{\lambda}
      g_{\nu \mu} - q_{\nu} g_{\lambda \mu}) \nonumber \\
&   & \hspace{20 mm} + C_{5}^{\it A}(q^{2}) g_{\mu \nu} + \frac{C_{6}^{\it A}
      (q^{2})}{M^{2}} q_{\mu} q_{\nu} \mbox{\LARGE \}} u({\bf p}).
\end{eqnarray}
This current corresponds to Eq.~\ref{eq:axcur2}, if we make the following
identifications
\begin{equation}
d_{4}(q^{2})= \frac{\sqrt{3}}{2} \frac{M}{M_{\Delta}} C_{3}^{\it A}(q^{2}) = 0.
\end{equation}
In our approach, the vanishing
of this form factor arises from restricting
ourselves solely to quarks bound in S-wave states. Several other
calculations are consistent with our result (see Schreiner \cite{21} for
an overview of some) though the reasoning might be different.

Having eliminated $d_{4}(q^{2})$ we find for the other form factors
\begin{eqnarray}
d_{1}(0) & = & \sqrt{3} C_{5}^{\it A}(0), \\
d_{2}(0) & = & \frac{\sqrt{3}}{2} C_{4}^{\it A}(0), \\
d_{3}(0) & = & \frac{\sqrt{3}}{2} ( 2 C_{6}^{\it A}(0) - C_{4}^{\it A}(0) ).
\end{eqnarray}


\begin{thebibliography}{999}
\bibitem{14} Extensions such as the cloudy bag model do not have this
restriction, but we are here using the naive constituent quark model.
\end{thebibliography}
\end{document}